\documentclass{aa}  
\usepackage{soul}
\usepackage{graphicx}
\usepackage{txfonts}
\begin{document} 

\title{The 2024 July 16 Solar Event: A Challenge To The Coronal Mass Ejection Origin Of Long-Duration Gamma-Ray Flares}
 
\author{
    Alessandro~Bruno\inst{1,2}
    \and
    Melissa~Pesce-Rollins\inst{3}
    \and
    Silvia~Dalla\inst{4}
    \and
    Nicola~Omodei\inst{5}
    \and
    Ian~G.~Richardson\inst{1,6}
    \and
    James~M.~Ryan\inst{7}
}

\institute{
Heliophysics Division, NASA Goddard Space Flight Center, Greenbelt, MD, USA\\
\email{alessandro.bruno-1@nasa.gov}
\and
Department of Physics, Catholic University of America, Washington, DC, USA
\and
Istituto Nazionale di Fisica Nucleare, Sezione di Pisa, I-56127 Pisa, Italy
\and
Jeremiah Horrocks Institute, University of Lancashire, Preston PR1 2HE, UK
\and
W.W. Hansen Experimental Physics Laboratory, Kavli Institute for Particle Astrophysics and Cosmology, Department of Physics and SLAC National Accelerator Laboratory, Stanford University, Stanford, CA 94305, USA
\and
Department of Astronomy, University of Maryland, College Park, MD, USA
\and
Space Science Center, University of New Hampshire, Durham, NH, USA
}

\date{October 30, 2025\\
\center{\large Accepted for publication in A\&A}}

\authorrunning{Bruno et al.}
\titlerunning{The 2024 July 16 Solar Event: A Challenge To The CME Origin Of LDGRFs}

  \abstract
  {
We present a multi-spacecraft analysis of the 2024 July 16 Long-Duration Gamma-Ray Flare (LDGRF) detected by the \textit{Large Area Telescope} on the \textit{Fermi} satellite. The measured $>$100 MeV $\gamma$-ray emission persisted for over seven hours after the flare impulsive phase, and was characterized by photon energies exceeding 1 GeV and a remarkably-hard parent-proton spectrum. In contrast, the phenomena related to the coronal mass ejection (CME)-driven shock linked to this eruption were modest, suggesting an inefficient proton acceleration unlikely to achieve the energies well-above the 300 MeV pion-production threshold to account for the observed $\gamma$-ray emission. Specifically, the CME was relatively slow ($\sim$600 km/s) and the accompanying interplanetary type-II/III radio bursts were faint and short-duration, unlike those typically detected during large events. In particular, the type-II emission did not extend to kHz frequencies and disappeared $\sim$5.5 hours prior to the LDGRF end time. Furthermore, the associated solar energetic particle (SEP) event was very weak, short-duration, and limited to a few tens of MeV, even at magnetically well-connected spacecraft. These findings demonstrate that a very-fast CME resulting in a high-energy SEP event is not a necessary condition for the occurrence of LDGRFs, challenging the idea that the high-energy $\gamma$-ray emission is produced by the back-precipitation of shock-accelerated ions into the solar surface. The alternative origin scenario based on local particle trapping and acceleration in large-scale coronal loops is instead favored by the observation of giant arch-like structures of hot plasma over the source region persisting for the entire duration of this LDGRF.
}

   \maketitle


\section{Introduction}\label{s:Introduction}
Long-duration $\gamma$-ray flares (LDGRFs) are a %
solar phenomenon often associated with large eruptions, characterized 
by photon emission %
extending up to the GeV range
and persisting for tens of minutes up to tens of hours 
after the flare impulsive phase, when other flare features (e.g., X rays) are absent or greatly suppressed 
(\citealp{ref:AKIMOV1991,ref:KANBACH1993,ref:AKIMOV1996,ref:RYAN2000,ref:RANK2001,ref:AJELLO2014}).
Launched in 2008 on board the \emph{Fermi} satellite, the \textit{Large Area Telescope} \citep[LAT;][]{LATPaper} has enabled a systematic investigation of these events, significantly improving the frequency and quality of available experimental observations (see \citealp{ref:AJELLO2021} and references therein). 
In particular, the \textit{first LAT solar-flare catalog} \citep{ref:AJELLO2021} contains detailed information about 45 high-energy $\gamma$-ray events (37 of which were LDGRFs), including
light-curves, energy spectra, inferred parent-proton numbers and spectral indices, as well as the emission-centroid localization for the brightest eruptions.
This unprecedented dataset offers a valuable foundation for statistical analyses and deeper investigation into the physical mechanisms driving $\gamma$-ray production in solar eruptive events.

While it is well established that $\gamma$ rays with energies greater than 50 MeV in LDGRFs are predominantly produced by the decay of neutral pions \citep{ref:RANK2001} created in the interactions of $>$300 MeV protons and $>$200 MeV/n $\alpha$ particles with the photosphere and chromosphere (e.g., \citealp{ref:VILMER2011}), the physical mechanism accelerating the parent ion population is still debated. The characteristics of the emission indicate an often time-extended source distinct from the impulsive flare. 
Moreover, the lack of features in the temporal profile on scales much shorter than the overall decay time suggests a continuous acceleration occurring over large volumes ($\mathcal{L}$ $\gtrsim$ 1 solar radii, R$_{s}$), smoothing over the details of the dynamics, followed by a slow and prolonged transport of accelerated particles to the solar surface to produce the $\gamma$ rays \citep{ref:RYANLEE1991,ref:RYAN2000,ref:RANK2001}.

Among the proposed origin scenarios, the one that received the most attention in the literature
consists of time-extended acceleration of ions at coronal mass ejection (CME)-driven shocks -- the dominant process responsible for gradual solar energetic particle (SEP) events observed in situ -- with subsequent back-propagation to the lower layers of the solar atmosphere
\citep{ref:WILD1963,ref:CLIVER1993,ref:KOCHAROV2015,ref:PESCE2015,ref:PLOTNIKOV2017,ref:JIN2018,ref:KAHLER2018,ref:SHARE2018,ref:GOPALSWAMY2018,ref:KOULOUMVAKOS2020,ref:PESCE2022}. 
This model 
was put forward because LDGRFs tend to be accompanied by fast CMEs associated with large SEP events, often with energies typical of ground-level enhancements (GLEs), suggesting that the ions producing the SEP events and the LDGRFs are part of the same population of energetic shock-accelerated particles.
However, this statistical association is not rigorous and contradicted by several counterexamples, including LDGRFs linked to slow/no CMEs, LDGRFs accompanied by small/no SEP event, and high-energy SEP events that are not associated with LDGRFs \citep{ref:BRUNO2023}. 

A major issue with the CME paradigm is the effect of magnetic mirroring, which strongly impedes the particle back-precipitation from the shock. Essentially, only ions injected nearly parallel to the coronal or interplanetary magnetic field (IMF) lines near the shock, in a narrow loss cone, can reach 
regions of the solar atmosphere that are sufficiently dense to
result in nuclear interactions that produce $\gamma$ rays \citep{ref:HUDSON2018,ref:KLEIN2018}.
The influence of mirroring is particularly relevant at late times in the photon emission, when the shock is rather distant from the Sun, as typically occurs %
in the longest-duration $\gamma$-ray events. 
Although relatively-good agreement with experimental data has been reported by some modeling studies based on the CME scenario, these are typically limited to the initial phase (first few hours), when the shock is still close to the Sun (e.g., \citealp{ref:JIN2018,ref:KOULOUMVAKOS2020}).
The mirroring problem was extensively investigated by \citet{ref:HUTCHINSON2022} 
by using %
3D test-particle simulations with varying turbulence levels.
They estimated that the fraction of shock-accelerated particles able to reach the solar surface is $\lesssim$2\%, and rapidly decreases with increasing heliocentric distance, exacerbating the problem for the fastest CMEs. Conversely, the precipitation fractions $N_{LDGRF}/(N_{LDGRF}+N_{SEP})$
required to explain both LDGRF and SEP data %
under the hypothesis of a common shock origin can be up to 20 times higher, even under conservative assumptions on the LDGRF, SEP-event, and interplanetary-scattering modeling. This was demonstrated by 
\citet{ref:DENOLFO2019} and \citet{ref:BRUNO2023}, %
who compared for 14 LDGRFs the numbers of $>$500 MeV protons producing the LDGRFs as inferred from the LAT observations ($N_{LDGRF}$) and the numbers of in-situ $>$500 MeV SEP protons based on multi-point measurements at 1 AU ($N_{SEP}$).
Such results challenge the current CME-origin models of LDGRFs
which assume
a common parent 
population accelerated by shocks, as they indicate that the measured SEPs are insufficient to account for the detected high-energy $\gamma$-ray emission.

The particle back-propagation is also disfavored by the presence of the CME structure following the shock (e.g., \citealp{ref:ZURBUCHEN2006}), which potentially complicates
the path of particles to %
the photosphere. 
In particular, the transport throughout the environment behind the shock ``nose'', where the highest-energy particles producing the LDGRFs are expected to be accelerated, is strongly limited by the sheath and compression-enhanced turbulence, making it difficult to achieve an efficient particle precipitation 
resulting in a remarkably-smooth decay of the $\gamma$-ray emission over multiple hours. 
It has been
proposed that the particle propagation from the shock nose might occur along open field lines passing through the sheath region (see Figure 14 in \citealp{ref:GOPALSWAMY2020}). In general, while this scenario might be plausible close to the Sun, where the shock shape is still dome-like, it ignores the downstream turbulence and the potential complexity of the open ambient coronal fields that the shock is moving through. In addition, it unjustifiably assumes that sheath field lines at the nose of the shock map back directly to the 
eruption site. In fact, as the CME moves further from the Sun, the shock becomes broader and does not wrap around its body; thus, 
the region at the Sun magnetically connected to the shock
is expected to be more extended and diffuse in longitude
(see, e.g., Figure 2 in \citealp{ref:ZURBUCHEN2006}).

The requirement of a prolonged acceleration of high-energy ions in the CME-origin scenario for LDGRFs 
is also challenged 
by the rapid decrease of the shock-acceleration efficiency with increasing heliospheric distance. In fact, protons with energies typical of GLEs, needed to account for the detected $>$100 MeV $\gamma$-ray emission,
are accelerated and released by a CME-driven shock within a few R$_{s}$ (see, e.g., \citealp{ref:ZANK2000,ref:REAMES2009,ref:GOPALSWAMY2013}).
In contrast, as mentioned above, the shock can be significantly far (from several tens to hundreds of R$_{s}$) from the Sun during the decay phase of LDGRFs,
so a dominant contribution from the precipitation of shock-accelerated particles appears improbable even when the discussed transport constraints are neglected.

As further support for the CME paradigm for the high-energy $\gamma$-ray emission,
\citet{ref:GOPALSWAMY2018,ref:GOPALSWAMY2019b} claimed a close statistical relationship between the duration of LDGRFs, and the duration and the end frequency of the concomitant interplanetary type-II radio emission. However, this result was questioned by \citet{ref:BRUNO2023}, who only found a poor/moderate correlation when using the LDGRF onset and end times obtained by \citet{ref:SHARE2018} and \citet{ref:AJELLO2021}, based on a much more rigorous approach to the analysis of the \textit{Fermi}-LAT data, and including the LDGRF events with a duration shorter than three hours, excluded by \citet{ref:GOPALSWAMY2018,ref:GOPALSWAMY2019b}. Indeed, a direct link between the two processes is not obvious, since the $\gamma$-ray and the type-II emissions are produced by different particles (ions vs. electrons) in distinct heliospheric regions (solar surface vs. interplanetary space). In particular, at the time when the type-II burst disappears, the shock is typically very far from the Sun, especially for fastest CMEs. For instance, for the 2012 January 23 and 2012 March 7 LDGRFs, the shocks were at 0.6 and 0.8 AU, respectively \citep{ref:MAKELA2023}. 
However, an efficient back-precipitation of high-energy ions from such large shock height is highly unlikely 
in light of the above discussion on the limitations affecting particle transport and acceleration in the CME-origin model.
In general, any apparent connection between LDGRFs and shock-related aspects, such as the interplanetary type-II burst duration, the CME speed/width and the SEP-event size, does not necessarily imply a causal relationship and might be just a manifestation of the so-called ``big-flare syndrome'' \citep{ref:KAHLER1982}: energetic phenomena are statistically more likely to occur together in large solar eruptions even when there is no common underlying physical mechanism.

An alternative to the CME scenario is provided by the trapping and acceleration of particles within large-scale coronal loops  \citep{ref:RYANLEE1991,ref:MANDZHAVIDZERAMATY1992,ref:AKIMOV1996,ref:RYAN2000,ref:CHUPPRYAN2009,ref:GRECHNEV2018,ref:DENOLFO2019,ref:BRUNO2023,ref:KOCHANOV2024}. They consist of quasi-static bipolar magnetic-flux structures filled with plasma and often appear during the gradual phase of two-ribbon flares and CME liftoff via field-line reconnection, creating a system of arches that can persist for several hours. In this case,
seed ions -- possibly flare-accelerated particles -- are injected into and trapped in such giant loops, where they are simultaneously accelerated by the 2$^{nd}$-order Fermi process up to GeV energies, before diffusing to the loop ends and precipitating into the solar surface leading to the high-energy $\gamma$-ray emission.
Under quasi-linear diffusion theory with realistic coronal turbulence levels, modeling of observed X-ray and $\gamma$-ray decay times requires coronal-loop lengths of order 1 R$_{s}$ \citep{ref:RYANLEE1991,ref:RYAN2000}. Such scales\footnote{While this $\sim$1 R$_{s}$ threshold serves only as a guideline, it strikes a balance between avoiding loop opening into the solar wind and requiring a realistic turbulence spectrum -- shorter loops would demand implausibly intense or steep turbulence to sustain multi-hour decay profiles.} provide the pitch-angle scattering and magnetic confinement needed for prolonged trapping of high-energy particles and their delayed precipitation.
Wave energy is continually provided from below, and magnetic turbulence or Alfv\'en waves with magnetic-field fluctuations having a relative amplitude of only $\delta B/B\sim$10\% are necessary for achieving efficient acceleration \citep{ref:RYANLEE1991,ref:RYAN2019}.
The loop model can easily reproduce the temporal evolution of LDGRFs, in particular the delayed onset due to the time required for the ion energies to exceed the pion-production threshold, and the prolonged emission. Because the loop scenario is local and the particles propagate diffusively, it can naturally produce smooth exponential decays in the $\gamma$-ray emission temporal profiles. %
In these loops no sequence of magnetic connections/disconnections would occur, as would be expected for a large-scale feature like a CME propagating in the interplanetary medium with rapidly-evolving connectivity to the Sun.  An excellent example is provided by 
the light curve of
the 2017 September 10 LDGRF \citep{ref:OMODEI2018,ref:RYAN2019}.
The emission decay time is predicted to depend on the turbulence level and loop size. %
Furthermore, turbulence in the loop -- the major factor accelerating particles to high energies -- is characterized by a predicted mean free path that is considerably smaller (a fraction of the loop length, i.e. diffusive transport), resulting in strong pitch-angle scattering of trapped ions into the loss cone.
Therefore, the effect of magnetic mirroring is significantly less relevant for the loop scenario, leading to efficient precipitation into the solar surface \citep{ref:RYANLEE1991,ref:RYAN2000}.

In addition, the coronal-loop scenario can straightforwardly explain the spatially-extended nature of the 
high-energy
$\gamma$-ray emission manifested in behind-the-limb events \citep{ref:ACKERMANN2017}, often used to support a CME-shock-related origin (e.g., \citealp{ref:CLIVER1993,ref:PLOTNIKOV2017,ref:JIN2018,ref:GOPALSWAMY2020}). It has been shown that giant coronal structures can in fact have widely-separated footpoints on the photosphere and, in particular, may connect a source region on the back side of the Sun to the visible part of the disk. %
For instance, using the \textit{Expanded Owens Valley Solar Array} microwave observations, \citet{ref:RYAN2019} identified the footpoints of a large coronal loop with a circular length of $\sim$1.4 R$_{s}$ during the 2017 September 10 western-limb eruption, associated with GLE 72. Similarly, using the gyro-synchrotron emission measured by the \textit{Nan\c{c}ay Radioheliograph} and the hard X-day data from the \textit{High Energy Neutron Detector} of the \textit{Gamma-Ray Spectrometer} on board the \textit{Mars Odyssey},
\citet{ref:GRECHNEV2018} concluded that the LAT emission during the 2014 September 1 backside event was generated by flare-accelerated particles trapped in static long coronal loops, that could be reaccelerated 
in these loops by a shock wave excited by the eruption, and not initially driven by a CME.
\citet{ref:KOCHANOV2024} confirmed the association of the 2014 September 1 LDGRF emission centroid with the bases of large-scale coronal loops connected to the flare site, supporting the flare origin of the high-energy protons injected and trapped into these giant arches. Another valuable example, although associated with a purely-impulsive $\gamma$-ray event, was provided by \citet{ref:PESCEROLLINS2024}, who analyzed the multi-wavelength observations during the 2022 September 29 behind-the-limb event, demonstrating that the $>$100 MeV $\gamma$-ray emission was associated with a magnetic structure connecting the parent active region on the far side of the Sun to the visible disk.

A more widespread acceptance of the loop model is primarily limited by the fact that, although these large-scale coronal magnetic structures are common, they are often difficult to visualize
due to insufficient soft X-ray and extreme-ultraviolet (EUV) emission measure, or because the electron energies and magnetic-field intensities in the loop do not lead to
detectable non-thermal (gyro)synchrotron emission.
Furthermore, the giant arches are typically obscured by the much brighter luminosity from the solar disk. In fact, all the aforementioned examples were characterized by a $\gamma$-ray emission
occurring 
near the solar limb. 
In addition, the loop model is affected by large uncertainties 
in the ambient conditions within the loops, and more theoretical work is necessary to evaluate the appropriate levels of turbulence and wave energy needed to achieve the required particle acceleration. 
Another unresolved question concerns the origin of the seed particle population injected into the coronal loops -- specifically, whether it arises from flare-associated processes or CME-driven shocks.
While shocks may be formed relatively close to the Sun (see, e.g., \citealp{ref:GOPALSWAMY2017}),
some time is required by the ions to be accelerated, and their release occurs at larger distances, often along open field lines. This geometry limits the likelihood of efficient injection into closed coronal loops, especially given the brief time spent by shocks near the loop system and the suppressive effects of magnetic mirroring. 
On the other hand, flare-accelerated particles originate in the low corona, where magnetic reconnection takes place. These regions are intrinsically linked to closed magnetic topologies, making injection into coronal loops geometrically and physically favorable.

In general, a hybrid scenario for the origin of LDGRFs might be plausible, given the diversity of observational signatures across events. In fact, partially due to the limitations in multi-messenger coverage, it is challenging to explain all reported cases with a single acceleration model.
It is likely that different mechanisms contribute to different events -- or even coexist within the same event -- with their relative importance governed by the specific physical conditions.
Nevertheless, based on the discussion above, it appears unlikely that CME-driven shocks can account for the $\gamma$-ray emission observed at late times during the longest-duration events.

In this work we present an analysis of the 2024 July 16 event, a remarkable LDGRF in terms of $\gamma$-ray energies and inferred parent-proton spectral hardness. We demonstrate that the concurrent shock-related phenomena contradict the presumed causal association between LDGRFs and SEPs, making this event an ideal counterexample to the CME paradigm. On the contrary, its interpretation within the loop model is favored by the observation of large-scale coronal structures persisting over the source region. This report is structured as follows.
In Section \ref{s:Observations and Analysis} we analyze the multi-spacecraft %
remote and in-situ observations collected during the event. We discuss their implications in terms of the two main LDGRF-origin scenarios in Section \ref{s:Discussion}. Finally, our summary and conclusions are given in Section \ref{s:Summary and Conclusions}.

\begin{figure}[!t]
    \centering
    \includegraphics[clip, trim=0 0 0 0,width=0.44\textwidth]{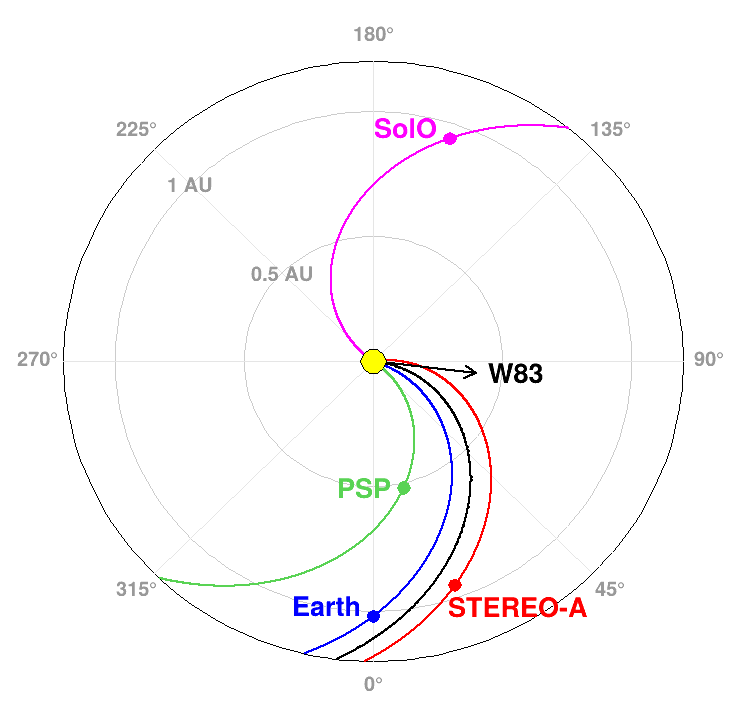}
    \caption{Spacecraft radial and longitudinal configuration %
    at 13:30 UT on 2024 July 16.}
    \label{fig:1}
\end{figure}
\begin{table*}[]
\centering
\begin{tabular}{c|l|l}
\hline
Time (UT) & Event & Mission/instrument \\
\hline\hline
13:10 & Onset hard X-ray emission & SolO/STIX \\
13:11 & Onset of the soft X-ray flare & GOES/XRS \\
13:20 & Onset of the type-III burst at 145--180 MHz & RSTN/SVTO \\
13:21 & End of the metric type-III burst & RSTN/SVTO \\
13:21 & Onset of the type-II burst at 25–-180 MHz & RSTN/SVTO \\
13:22$^{a}$ & Peak of the hard X-ray emission & SolO/STIX \\
13:26 & Peak of the soft X-ray flare & GOES/XRS \\
13:29$^{b}$ & End of the hard X-ray flare & SolO/STIX \\
13:30$^{c}$ & Appearance of giant coronal loops & SDO/AIA, GOES/SUVI \\
13:35 & Onset of the SEP proton event & SOHO/COSTEP \\
13:36$^{b}$ & End of the soft X-ray flare & GOES/XRS \\
13:36 & First appearance of the CME in the C2 FoV & SOHO/LASCO \\
13:36 & Onset of the DH type-III burst & \textit{Wind}/WAVES \\
13:43 & Moderate type-IV burst at 25--83 MHz & RSTN/SVTO \\
13:44 & Onset of the DH type-II burst & STEREO/WAVES \\
13:55 & End of the metric type-II burst & RSTN/SVTO \\
14:00 & Start of the first interval with LDGRF detection & \textit{Fermi}/LAT\\
14:23 & Highest-energy (1.68 GeV) photon detection & \textit{Fermi}/LAT\\
14:51 & End of the metric type-IV burst & RSTN/SVTO \\
15:00 & End of the DH type-III burst & \textit{Wind}/WAVES \\
15:00$^{d}$ & Onset of the SEP proton event & PSP/EPI \\
15:47 & End of the DH type-II burst & STEREO/WAVES \\
19:18 & Last measurement of the CME in the C3 FoV & SOHO/LASCO \\
20:38 & End of the last interval with LDGRF detection & \textit{Fermi}/LAT\\
21:15 & Estimated end time of LDGRF & \textit{Fermi}/LAT\\
$^{+1}$02:00$^{c}$ & Disappearance of giant coronal loops & SDO/AIA, GOES/SUVI \\
\hline
\end{tabular}
\\
$^{a}$Based on a visual inspection of interpolated intensities (see the text for details). $^{b}$Estimated as the time when the X-ray intensity decays to a level halfway between the peak value and the pre-flare background level. $^{c}$Rough estimate. $^{d}$Based on a visual inspection of the 30 MeV proton-intensity profile, and accounting for the $\sim$24-min travel delay between the PSP location ($\sim$0.5 AU) and 1 AU.
\caption{Timeline of the 2024 July 16 eruption.}
\label{tab:Timeline}
\end{table*}

\section{Observations and Analysis}\label{s:Observations and Analysis}
The 2024 July 16 solar event was well observed by several instruments at different heliospheric locations, including the Earth / L1 Lagrange point, \textit{Solar TErrestrial RElations Observatory-A} (STEREO-A), \textit{Parker Solar Probe} (PSP) and \textit{Solar Orbiter} (SolO). The corresponding radial/longitudinal configuration in Stonyhurst coordinates 
is shown in Figure \ref{fig:1}. 
The solid curves denote the IMF lines passing through
each observer, based on a simple Parker-spiral model\footnote{See Appendix in \citet{ref:BRUNORICHARDSON2021}.} and the locally-measured solar-wind speeds. The dashed curve marks the IMF line connected to the flare site, whose longitude (see Section \ref{s:XRay Emission}) is indicated by the arrow. 
An approximate timeline of the different aspects characterizing the solar event, presented in the next sections, is summarized in Table \ref{tab:Timeline}.

\begin{figure*}[h!]
    \centering
    \includegraphics[width=1\textwidth]{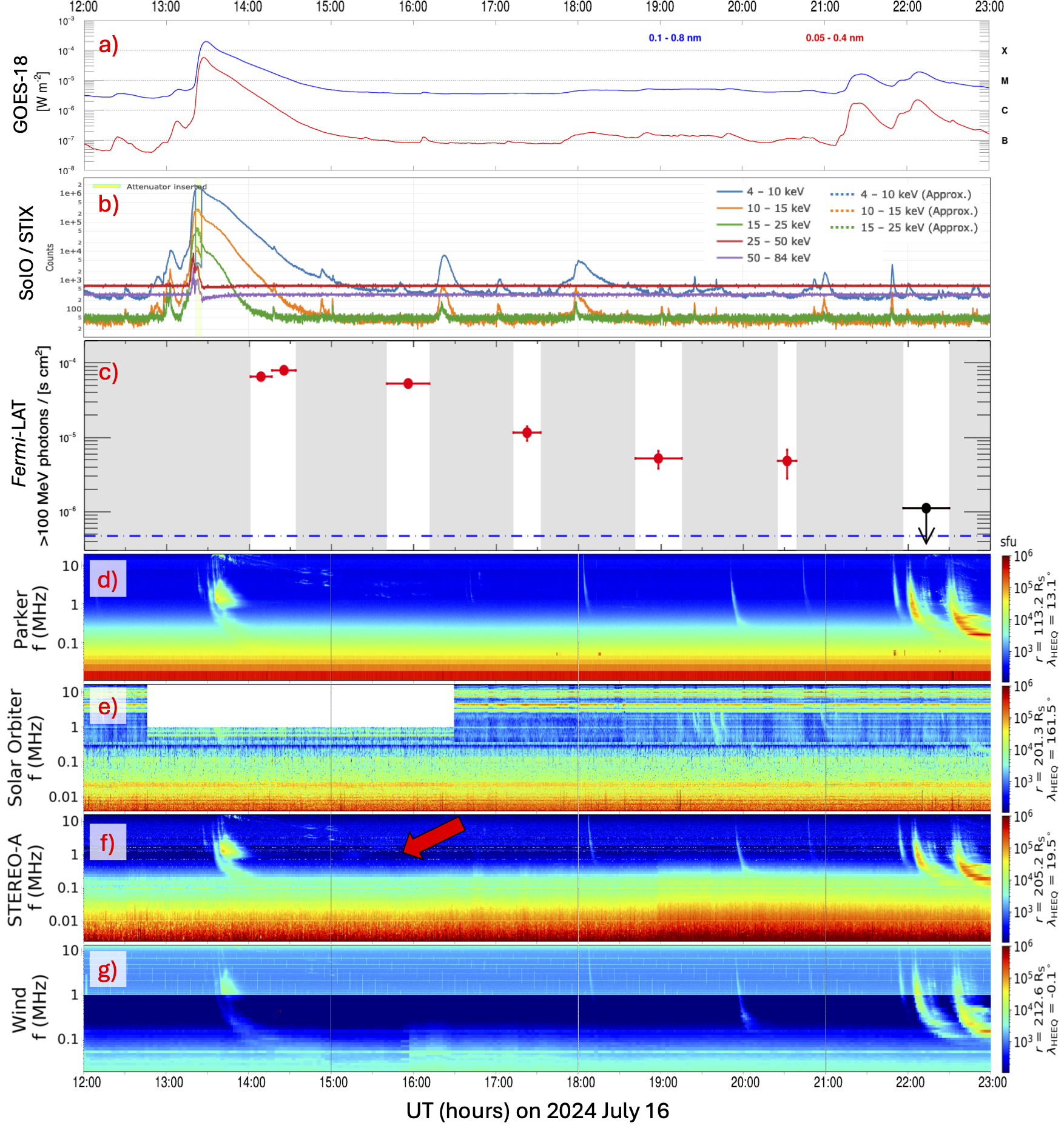}
    \caption{Remote-sensing observations between 12--23 UT on 2024 July 16.
    From top to bottom: the soft X-ray flux measured by GOES-18 (a); the hard X-ray counts measured by SolO (b); the $>$100 MeV $\gamma$-ray flux measured by \textit{Fermi}-LAT (c), with the blue dashed line marking the quiet-Sun background and the shaded areas indicating the intervals where the Sun was outside the detector FoV; and the radio emission measured by PSP (d), SolO (e), STEREO-A (f) and Wind (g). The red arrow in panel f) marks the end of the interplanetary type-II emission at 15:47 UT (1.4 MHz) according to the CDAW DH type-II burst catalog.}
    \label{fig:2}
\end{figure*}

\subsection{X-Ray Emission}\label{s:XRay Emission}
The parent flare originated in Active Region 13738 near the western limb of the Sun, with an S05W83 position estimated by \textit{SolarSoft}\footnote{\url{http://www.lmsal.com/solarsoft/}}. The soft X-ray irradiance measured by the \textit{X-Ray Sensor} (XRS) on board the \textit{Geostationary Operational Environmental Satellite}-18 (GOES-18) at short (0.05--0.4 nm) and long (0.8--1.0 nm) wavelength bands is shown in Figure \ref{fig:2}a. The flare started at 13:11 UT and reached a peak corresponding to an X1.9 class at 13:26 UT, ending ten minutes later (13:36 UT).

No hard X-ray data from the \textit{Gamma-Ray Burst Monitor} (GBM) on board \textit{Fermi} are available during the event impulsive phase until 13:48 UT.
However,
the flare was directly observed by SolO, 
located on the far side of the Sun relative to Earth. Specifically, Figure \ref{fig:2}b displays the unnormalized hard X-ray counts measured by the \textit{Spectrometer Telescope for Imaging X-rays} (STIX), in different energy intervals between 4--84 keV.
The yellow-shaded area marks the approximated count values 
around the peak derived during the interval when the instrument attenuator was inserted ($\sim$13:21--13:25 UT). Compared to GOES, the SolO data exhibit additional peaks linked to C-class eruptions.

\begin{table*}[ht!]
    \centering
    \begin{tabular}{c|c|c|c|c|c}
    \hline
        Time interval & Intensity & Model & Index & Cutoff Energy & Proton index \\
        (UT) & (10$^{-5}$ cm$^{-2}$ s$^{-1}$) &  &  & (MeV) & $\alpha_{p}$ \\
        \hline
        \hline
        14:00:57 -- 14:16:29 & $ 6.5 \pm 0.8 $  & PLEXP & $ -1.2 \pm 0.5 $  & $ \left(3.2 \pm 1.4\right) \times 10^{2} $ & $ 3.46 \pm 0.28 $ \\
        14:16:29 -- 14:34:08 & $ 7.9 \pm 0.8 $ & PLEXP & $ -0.5 \pm 0.4 $  & $ \left(2.2 \pm 0.6\right) \times 10^{2} $ & $ 3.31 \pm 0.21 $ \\
        15:40:18 -- 16:11:24 & $ 5.3 \pm 0.4 $  & PLEXP & $ 0.30 \pm 0.13 $ & $ \left(1.0 \pm 0.1\right)\times 10^{2} $ & $ 4.26 \pm 0.29 $\\
        17:12:38 -- 17:32:17 & $ 0.96 \pm 0.22 $ & PL  & $ -2.46 \pm 0.25 $ & ... & ...\\
        18:41:22 -- 19:15:13 & $ 0.52 \pm 0.14 $ &PL & $ -2.49 \pm 0.29 $  & ... & ...\\
        20:25:06 -- 20:39:30 & $ 0.36 \pm 0.16 $  & PL & $ -2.7 \pm 0.6 $  & ... & ...\\
        21:56:40 -- 22:30:02 & $ <0.1$ & PL & ... & ... & ...\\
        \hline
    \end{tabular}
    \caption{Results for the time-resolved likelihood analysis of the LAT data. The $>$100 MeV $\gamma$-ray flux intensity, spectral parameters (power-law indices and cutoff energies) of the best-fitting photon model (reported in column 3), and the inferred parent-proton spectral indices are listed. See the text for details.}
    \label{tab:best_proton_index}
\end{table*}

\subsection{CME Properties}\label{s:CME Properties}
The associated CME appeared in field of view (FoV) of the \textit{Large Angle and Spectrometric Coronagraph} (LASCO) C2 on board the \textit{SOlar and Heliospheric Observatory} (SOHO) at 13:36 UT.
It was classified as poor-event, partial-halo (236$^{\circ}$ angular width) CME in the \textit{Coordinated Data Analysis Workshop}\footnote{\url{https://cdaw.gsfc.nasa.gov/CME_list/}} (CDAW) catalog. Its sky-plane (projected) velocity slowly decreased with a -7.82 m/s$^{2}$ acceleration from $\sim$650 km/s at a height of 2.45 %
R$_{s}$, to $\sim$500 km/s at 20.42 R$_{s}$ before exiting the C3 coronagraph FoV at 19:18 UT. %
The corresponding average speed was 580 km/s; the de-projected (3D or space) value, not provided by CDAW, is expected to be similar given the proximity to the solar limb.
According\footnote{See \citet{ref:RICHARDSON2015} for a comparison of the CME properties in different catalogs.} to the \textit{Space Weather Database Of Notifications, Knowledge, Information}\footnote{\url{https://ccmc.gsfc.nasa.gov/donki/}} (DONKI), the CME had a space speed of 642 km/s and a half width of 38$^{\circ}$, based on LASCO data up to 12.2 R$_{s}$, and propagated very close to the ecliptic plane with a N07W80 direction, accounting for the favorable ($\sim$4$^{\circ}$) solar $B_{0}$ angle (the heliographic latitude of the Sun's center).

The LASCO C2/C3 velocity estimates are in agreement with the mean value ($\sim$570 km/s between $\sim$5.5--14.5 R$_{s}$) obtained by using the CME data from the COR2 coronagraph on board STEREO-A.
We also used images from the STEREO-A COR1 and \textit{Extreme UltraViolet Imager} (EUVI) instruments %
to investigate the CME speed closer to the Sun. %
The derived value %
($\sim$710 km/s between $\sim$1.2--3 R$_{s}$)
is consistent with the low-altitude extrapolation of the C2/C3 and COR2 measurements. This rules out any possibility that the CME %
linked to the 2024 July 16 event
had a much higher speed in the low corona, and might have generated a strong shock capable of efficient particle acceleration.

Finally, the next CME in the CDAW/DONKI catalogs was observed around 18 UT, with similar speed but a south-eastward traveling direction (-57$^{\circ}$ latitudinal angle in DONKI), thus did not interfere with the event under investigation.

\subsection{Gamma-Ray Emission}\label{s:Gamma-Ray Emission}
$\gamma$-ray emission at energies greater than 100 MeV was observed by the \emph{Fermi}-LAT 
\citep{LATPaper} from this event starting shortly after 14:00 UT, when the Sun came into its FoV. We selected the Pass 8~\citep{atwood2013pass8realizationfermilat,bruel2018fermilatimprovedpass8event} Source class LAT data (using \texttt{P8R3$\_$SOURCE$\_$V3} instrument response functions) from a 10$^{\circ}$ circular region centered on the Sun and within 100$^{\circ}$ from the local zenith (to reduce contamination from Earth's limb) and fit the data with three different models. The first two are a simple power-law (PL) and a power-law with an exponential cut-off (PLEXP), phenomenological functions\footnote{The definition of the models used can be found here: \url{https://fermi.gsfc.nasa.gov/ssc/data/analysis/scitools/source_models.html}}  that may describe bremsstrahlung emission from relativistic electrons. 
The third model is a series of templates based on a detailed study of the $\gamma$ rays produced from the decay of pions originating from accelerated protons 
(for more details, see \citealp{murp87, Murphy_2009}).
We used the Multi-Mission Maximum Likelihood (\texttt{3ML})\footnote{\url{https://threeml.readthedocs.io/en/stable/index.html}} framework with \texttt{fermitools}\footnote{\url{https://github.com/fermi-lat/Fermitools-conda/wiki}} version 2.0.8 to perform an unbinned likelihood analysis of the \emph{Fermi}-LAT data. 

The best-fit flux results for each time interval are shown in Figure \ref{fig:2}c, with the dashed horizontal line marking the average quiet-Sun photon intensity (4.7$\times$10$^{-7}$ cm$^{-2}$ s$^{-1}$) reported in \citet{ref:AJELLO2021}.
The $\gamma$-ray flux remained significant for more than 6 hours, with the highest value registered between 14:16--14:34 UT with an intensity of 7.9$\pm$ 0.8$\times$10$^{-5}$ cm$^{-2}$ s$^{-1}$. In particular, four photons with energies greater than 1 GeV were detected, clustering between 14:11 and 14:23 UT, with the highest energy photon of 1.68 GeV arriving at 14:23 UT. 
The $\gamma$-ray intensity dropped by more than an order of magnitude in the following intervals, with a statistically-significant signal until 20:38 UT. 
The Sun was again in the LAT FoV at 21:52 UT, with an upper-limit of 1.12$\times$10$^{-6}$ cm$^{-2}$ s$^{-1}$, indicated by the black point in Figure \ref{fig:2}c. By taking the midpoint between these two time bins, we estimated an end time occurring at 21:15 UT, implying a total LDGRF duration of 7.25 hours based on the first detection interval starting at 14 UT. 
Unfortunately, the Sun was in the LAT FoV for only $\sim$33 minutes between 13:00 and 15:00 UT, therefore we have no coverage of the prompt phase of the event, and the time in which the delayed phase started and peaked is uncertain.

During the first $\sim$2 hours (first three points in Figure \ref{fig:2}c), the statistics are sufficient to perform a time-resolved spectral analysis, and we found that the PLEXP model best describes the data at the 4$\sigma$ level\footnote{The likelihood ratio test and the associated test statistic TS \citep{Mattox:96} provide an estimate of the significance of the detection and the significance in $\sigma$ can be roughly approximated as $\sqrt{\rm TS}$.}.
Given that the curved model is preferred, we also fit the data with the pion templates described above to derive the power-law spectrum ($\propto E^{-\alpha_{p}}$) associated with the parent ion population that best reproduces the observations. In the remaining time intervals starting at 17:12 UT, the data is best fit by a simple power-law model.
The results of the analysis are summarized in Table~\ref{tab:best_proton_index}: the first two columns list the time intervals and the corresponding $>$100 MeV $\gamma$-ray intensity; the next three columns report the model used to fit the data and the relative parameters; finally, the last column shows the best proton index $\alpha_{p}$ obtained for the first three time bins with the pion templates.

\subsection{Radio Emission}\label{s:Radio Emission}
The radio emission during this event was detected by several space- and ground-based instruments, including the \textit{Radio and Plasma Wave experiment} (WAVES) on board \textit{Wind} and STEREO-A \citep{ref:BOUGERET1995, ref:BOUGERET2008}, FIELDS on board \textit{Parker Solar Probe} \citep{ref:BALE2016}
and the \textit{Radio Solar Telescope Network} (RSTN).
Unfortunately, the $>$1 MHz data from the \textit{Radio and Plasma Wave} (RPW; \citealp{ref:MAKSIMOVIC2020}) detector on board SolO from the far side of the Sun are affected by a large gap between $\sim$12:30--16:30 UT (see Figure \ref{fig:2}e).
A summary of spacecraft radio observations is displayed in panels d-g of Figure \ref{fig:2}, obtained from the \textit{Coordinated Radiodiagnostics Of CMEs and Solar flares}\footnote{https://parker.gsfc.nasa.gov/crocs.html} (CROCS).

\begin{figure*}[h!]
    \centering
    \includegraphics[width=0.75\textwidth]{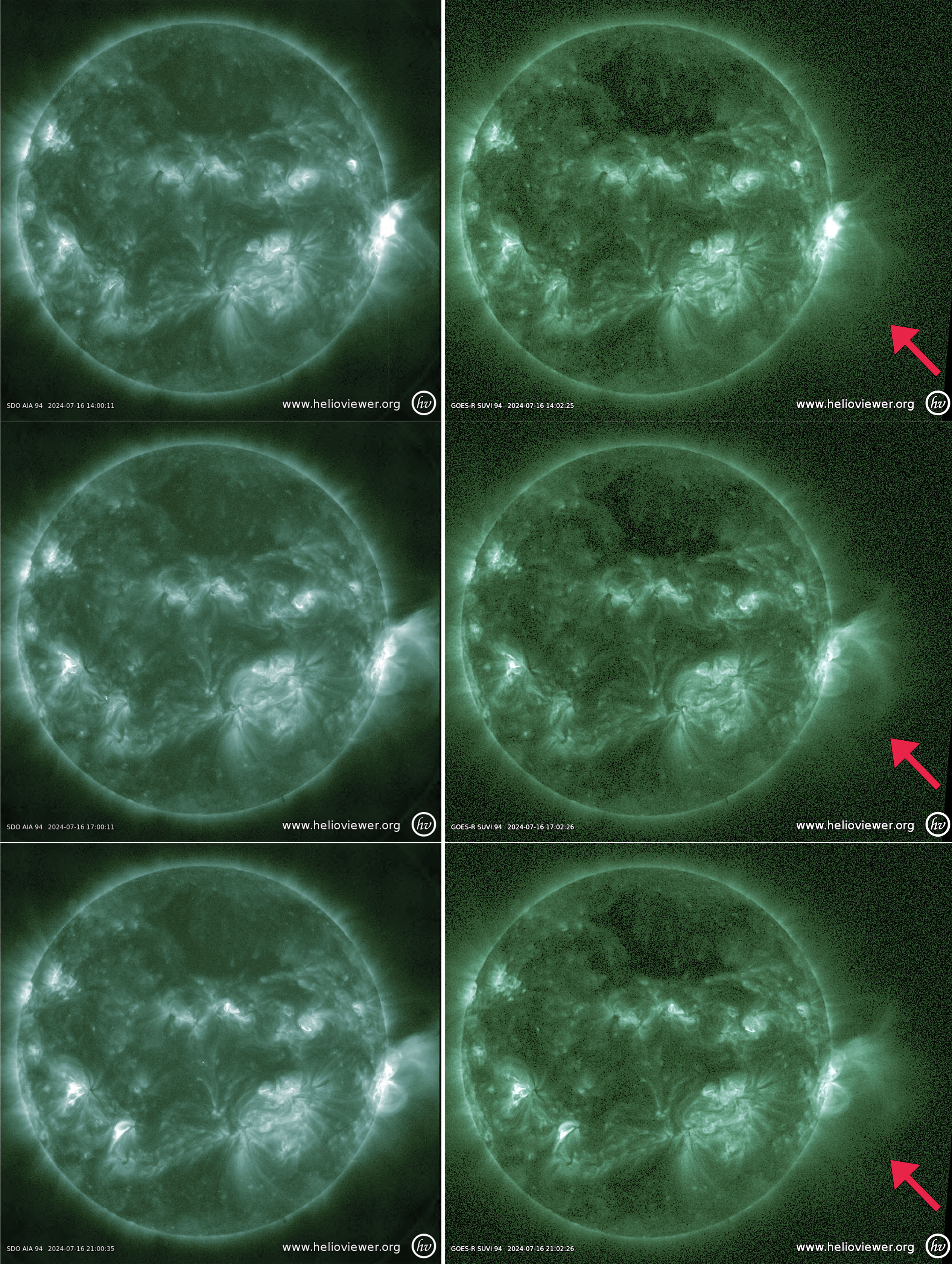}\\
    \caption{Extreme-ultraviolet images at 94 \r{A} taken by SDO/AIA (left) and GOES/SUVI (right) on 2024 July 16 at $\sim$14 UT (top panels), $\sim$17 UT (middle panels) and $\sim$21 UT (bottom panels), approximately corresponding to the start, middle and end times of the $>$100 MeV $\gamma$-ray emission (see Figure \ref{fig:2}c). The red arrows indicate the system of giant loops persisting over the source region for the entire LDGRF duration. 
    }
    \label{fig:3}
\end{figure*}

As listed in the NOAA Solar and Geophysical event reports\footnote{\url{https://www.swpc.noaa.gov/products/solar-and-geophysical-event-reports}.}, the RSTN \textit{San Vito Solar Observatory} (SVTO) station (sensitive to the metric wavelength spectrum) recorded a 
type-III radio burst in the 145--180 MHz range between 13:20--13:21 UT. Type-III radio emission was 
subsequently observed in the decameter-hectometric (DH) wavelengths by WAVES and FIELDS.
RSTN also reported a major type-II burst in the 25--180 MHz range between 13:21--13:55 UT, and a moderate type-IV burst between 13:43--14:51 UT in the 25--83 MHz range.
According to the CDAW DH type-II burst catalog\footnote{\url{https://cdaw.gsfc.nasa.gov/CME_list/radio/waves_type2.html}}, based on combined \textit{Wind} and STEREO observations, the $<$16 MHz emission (evident in the PSP and STEREO observations) 
only persisted for about two hours (13:44--15:47 UT) and did not extend below 1.4 MHz (see arrow in Figure \ref{fig:2}f). This feeble burst 
suggests that weak shock acceleration was limited to distances of a few tens of R$_{s}$,
based on the anticorrelation between the plasma frequency and the radial distance.
In contrast, the spectrum of the type-II emission associated with fastest and widest CMEs resulting in large SEP events typically extends to kilometric wavelengths %
(see, e.g., \citealp{ref:TEKLU2025} and references therein), hence to much larger heliodistances (up to a few hundreds of R$_{s}$).

\begin{figure*}[!t]
    \centering
    \includegraphics[width=0.75\textwidth]{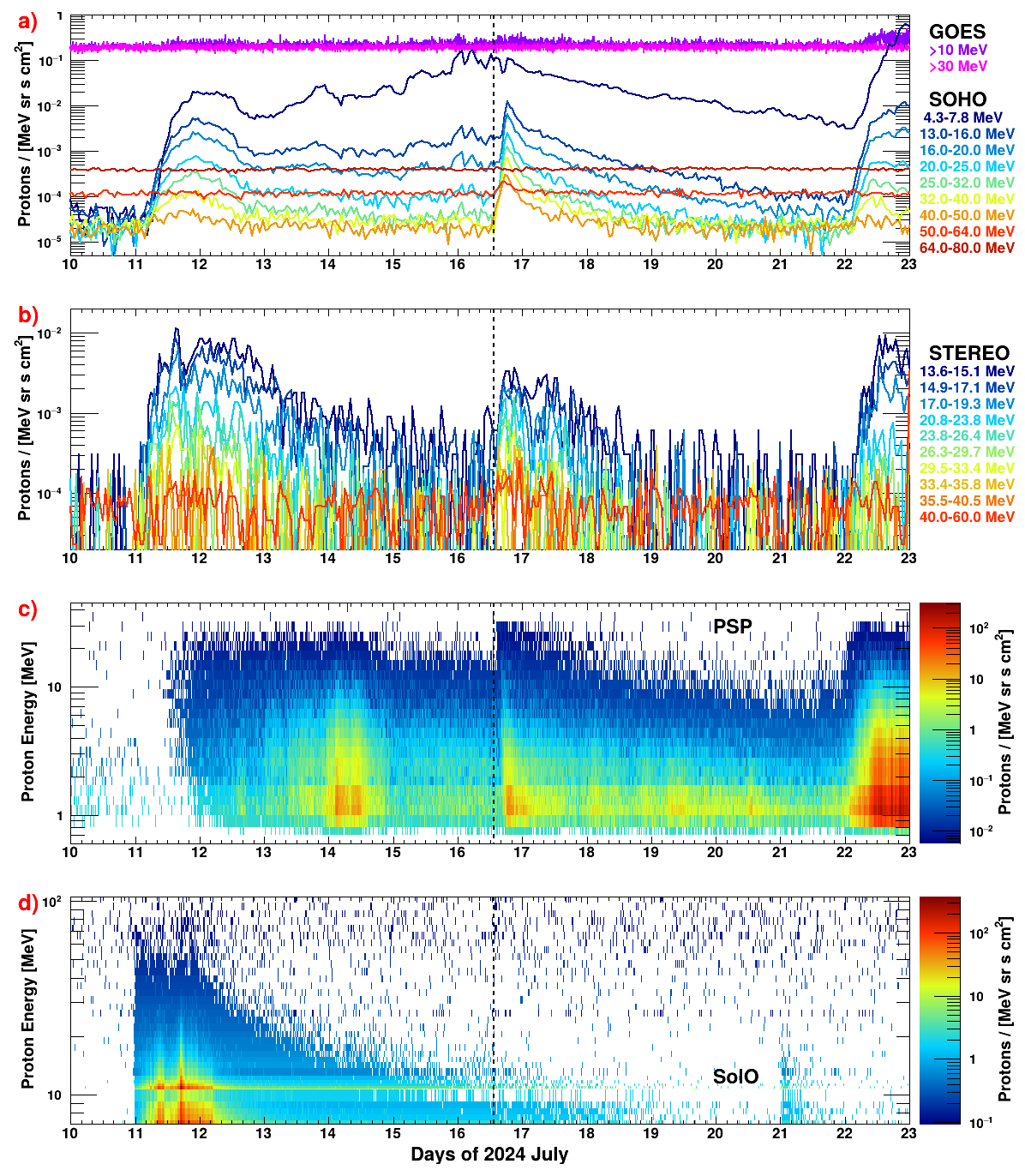}
    \caption{Temporal profiles of SEP intensities measured between 2024 July 10--23 by GOES and SOHO (a), STEREO-A (b), PSP (c) and SolO (d). The vertical dotted line marks the peak time of the X1.9-class flare associated with the LDGRF event.}
    \label{fig:4}
\end{figure*}
\subsection{Extreme Ultraviolet Emission}\label{s:Ultraviolet Emission}
Figure \ref{fig:3} 
displays the EUV wavelength images at 94 \r{A} taken by the \textit{Atmospheric Imaging Assembly} (AIA) on board the \textit{Solar Dynamics Observatory} (SDO) and the \textit{Solar Ultraviolet Imager} (SUVI) on board GOES at 14, 17, and 21 UT on 2024 July 16. While the former has higher resolution, the latter offers a larger FoV. Both instruments show the formation of long-lived coronal loops emerging right after the eruption onset, and persisting for several hours until $\sim$2 UT on July 17, i.e. well after the end of the LDGRF (see Table \ref{tab:Timeline}). Along with an arcade of relatively-small bright loops above the flare site, we note 
faint emission from
a system of
quasi-static giant loops, indicated by the red arrows, with an estimated circular length of $\sim$1.1--1.4 R$_{s}$, and a $\sim$0.4 R$_{s}$ height. %

Unlike the small-scale loops, these large structures are clearly visible only 
in the 94 \r{A} channel, corresponding to plasma temperatures of $\sim$6--10 MK.
This rules out typical post-flare cooling behavior, where loops transition through multiple EUV channels. Instead, it supports a scenario of localized, sustained heating not driven by impulsive flare energy release, but likely associated with the trapping of an energetic particle population \citep{ref:BROOKS2011,ref:DELZANNA2018}. 
The persistence of the EUV emission throughout the entire duration of the LDGRF further suggests continuous energy input into the loop system. This may arise from trapped particles undergoing 2$^{nd}$-order Fermi acceleration (e.g., \citealp{ref:PETROSIAN2012}) and turbulent wave activity (such as Alfv\'en or compressive magnetosonic modes) interacting with both the trapped particles and the plasma in the loops (e.g., \citealp{ref:VANBALLEGOOIJEN2017,ref:CALLINGHAM2024}), although the precise energy-transfer pathways remain uncertain.

As discussed in Section \ref{s:Introduction} for other events, the visibility of the giant loops is highly favored by the limb location, far from the brightness of the disk.
We also note that
the EUV images at various wavelengths do not show any evidence 
for a wave propagating away from the solar flare which suggests that a fast shock was absent during this event.

\subsection{Solar Energetic Particles}\label{s:Solar Energetic Particles}

The proton intensities measured between 2024 July 10--23 by near-Earth detectors, SolO and PSP are summarized in Figure \ref{fig:4}. As displayed in Figure \ref{fig:1}, the Earth
was magnetically well connected to the source region of the 2024 July 16 LDRGF, its footpoints being located at N04W72. Specifically, by interpolating the solar-wind speed value (330 km/s) between the \textit{Wind} (based on OmniWeb data\footnote{\url{https://omniweb.gsfc.nasa.gov/}}) and STEREO-A measurements, the footpoint longitudinal separation $\Delta Lon$ with respect to the nominal flare location was about -11$^{\circ}$ (East of the flare), while the latitudinal deviation $\Delta Lat$ was about +9$^{\circ}$ (North of the flare). %
The magnetic connection is even improved ($\Delta Lon \sim$ -8$^{\circ}$ and $\Delta Lat$ $\sim$ -3$^{\circ}$) when the CME direction from DONKI (N07W80) is used. 
Consequently, near-Earth detectors would be expected to measure SEPs 
accelerated by the strongest regions of the shock close to its nose. 
However, despite the favorable viewpoint, the proton signal is barely visible only in the GOES-18 10-min resolution intensities integrated above 10 MeV reported in Figure \ref{fig:4}a, where the dashed vertical line indicates the peak time of the X1.9 flare on 2024 July 16. In general, the SEP intensities at Earth remained a least a factor $\sim$20 lower than the operational threshold values used by the NOAA Space Weather Prediction Center to trigger warnings\footnote{see \url{https://www.swpc.noaa.gov/products/goes-proton-flux}.}, so no SEP event was reported.
Figure \ref{fig:4}a also displays the hourly proton measurements from the SOHO/\-\textit{Comprehensive Suprathermal and Energetic Particle Analyzer} (COSTEP; \citealp{ref:MULLERMELLIN1995}) detector between 4.3 and 80 MeV, characterized by a significantly-lower instrumental background with respect to GOES particle sensors (see, e.g., \citealp{ref:RICHARDSON2023}). 
Here a modest SEP event with rapid rise and decay can be seen on July 16 at energies lower than 50 MeV.
The intensities below $\sim$30 MeV 
show an elevated background following an event (most clearly observed by SolO) associated with a far-side halo CME late on July 10.
The near-Earth SEP data for the July 16 event are 
consistent with relatively weak particle acceleration at a slow CME-driven shock.

STEREO A was also similarly well-connected to the eruption site, with a field line footpoint east of the flare with longitudinal and latitudinal separations of +13$^{\circ}$ and +11$^{\circ}$, respectively.
The hourly proton intensities
measured by the \textit{High Energy Telescope} \citep{ref:VONROSENVINGE2008}
are reported in Figure \ref{fig:4}b. Also in this case, the recorded SEP event was definitely weak, with a statistically-significant proton increase limited to channels below $\sim$35--40 MeV, and a soft event-integrated spectrum which can be described by a power-law fit $dN/dE\propto E^{-\alpha}$ with $\alpha$$\sim$4.6.

Figure \ref{fig:4}c displays the proton intensities measured by 
the \textit{Integrated Science Investigation of the Sun} (IS$\odot$IS) \textit{Energetic Particle Instruments} (EPI; \citealp{ref:MCCOMAS2016})
on board PSP, located at $\sim$0.52 AU radial distance. The spacecraft was connected to the eastern side of the flare, with footpoints located $\Delta Lon \sim$ -29$^{\circ}$ away from the eruption site. Similarly to the aforementioned observations, the recorded proton enhancement following the July 16 flare and CME does not extend above a few tens of MeV. 

Finally, Figure \ref{fig:4}d shows the data from 
the \textit{Energetic Particle Detector} and the \textit{High-Energy Telescope} on board 
SolO \citep{ref:PACHECO2020}, located at 0.94 AU on the far side of the Sun. The proton intensities were relatively high 
near the beginning of the interval shown 
due to an ongoing SEP event associated with eruption occurring late on July 11, and no increase possibly linked to the July 16 event is visible. 
The absence of any SEPs from the July 16 event is consistent with the slow CME and large ($\Delta Lon \sim$ +209$^{\circ}$) separation between the SolO field-line footpoint and the flare location (see Figure \ref{fig:1}).

\section{Discussion}\label{s:Discussion}  
The 2024 July 16 LDGRF measured by \textit{Fermi}-LAT, characterized by a $>$7 hour duration, was a remarkable event in terms of photon energies (up to 1.68 GeV). Moreover, when compared to all LDGRFs previously reported in \citet{ref:AJELLO2021}, the parent-proton spectrum %
inferred from the $\gamma$-ray observations is significantly hard, being characterized by 
a $\alpha_{p}^{min}$ = 3.31 $\pm$ 0.21
minimum proton-index value (see Table \ref{tab:best_proton_index}). %
Only the production spectrum of the exceptional LDGRF on 2017 September 10 \citep{ref:OMODEI2018} was harder,
with $\alpha_{p}^{min}$ =
3.30 $\pm$ 0.06, and
even major LDGRFs such as the 2012 March 7, the 2014 February 25 and the 2014 September 1 events exhibited a softer production spectrum ($\alpha_{p}^{min}$ $\simeq$ 3.72--3.78).
For comparison, 
the average minimum proton index based on the delayed-emission events analyzed in \citet{ref:AJELLO2021} is $\sim$5. 
We note that, as discussed by \citet{ref:SHARE2018}, the atmospheric attenuation associated with a near-limb source region might have hardened the $\gamma$-ray emission spectrum with respect to eruptions located at the disk center. However, this aspect is not expected to significantly affect our conclusions, as the inferred spectral variation is relatively small and comparable with the fit-parameter uncertainties. Furthermore, the same considerations apply to 
a number of LDGRFs in the LAT catalog occurring at the solar limb -- and associated with a considerably faster CME -- including the aforementioned 2014 February 25 and 2017 September 10 events.
Clearly,
the LAT observations for the 2024 July 16 LDGRF
imply the existence of an 
ion population
interacting at the Sun,
with energies %
much higher than the 300 MeV pion-production threshold. 

However,
the remote-sensing and in-situ measurements presented in the previous section for the 2024 July 16 event indicate that 
it is highly unlikely that
these energetic particles %
were
accelerated by the CME-driven shock.
This deduction is primarily supported by the 
following  three aspects.
\begin{itemize}
\item The CME %
was modest in speed ($\sim$600 km/s) and width (partial halo in the CDAW catalog), slightly decelerating throughout the LASCO coronagraph FoV. Its kinematics clearly demonstrate that a very fast and wide CME with impulsive acceleration is not a necessary condition for the production of LDGRFs. However, this finding is not unusual: LDGRFs associated with CMEs with similar speeds include those on 2011 September 6 (650 km/s based on DONKI; 830 km/s based on CDAW), 2011 September 7 (750 km/s based on DONKI; partial halo, 575 km/s sky-plane based on CDAW), 2012 June 3 (absent in DONKI; partial-halo 605 km/s based on CDAW) and 2013 October 25 (477 km/s based on DONKI; partial-halo 587 km/s based on CDAW).  
When compared to the 2083 km s$^{-1}$ average velocity estimated for CMEs associated with the 23$^{rd}$ solar-cycle GLEs \citep{ref:GOPALSWAMY2012}, 
the shocks driven by these relatively-slow CMEs do not seem capable of efficiently accelerating ions to the high energies required to account for the detected $>100$ MeV $\gamma$-ray emission. For the 2024 July 16 event, this is corroborated by the lack of a kilometric type-II burst and confirmed by the low SEP intensities (see below). %
Other noteworthy events challenging the shock paradigm include the 2012 October 23 and No\-vem\-ber 27 LDGRFs, that were not even accompanied by a CME \citep{ref:SHARE2018,ref:AJELLO2021}. 
However,
with its $>$7 hours of emission, the 2024 July 16 LDGRF is unique with respect to those events linked to only a slow or no CME, which were typically characterized by a much shorter duration ($\lesssim$2 hours).

\item %
Although the source active region was magnetically well connected to the Earth and STE\-REO-A, the detected SEP event was of modest intensity, short duration and limited to energies below a few tens of MeV. The proton signal barely exceeded the instrumental background of the GOES satellites even in the $>$10 MeV channel.
The short ($\sim$2 days) duration of the SEP event is more consistent with an ``impulsive'' event than a large gradual event dominated by acceleration at a fast CME-driven shock, although no spacecraft was in a position to detect in situ any interplanetary shock that might have been associated with this event and evaluate its properties.
The type-III emission was also weak and of short duration, in contrast to that associated with major SEP events (e.g., \citealp{ref:CANE2002}).
Overall, these observations appear to be
inconsistent with the CME-origin scenario of LDGRFs, given the 
lack of an intense population of shock-accelerated protons,
even at energies around an order of magnitude below the 
pion-production threshold,
required to produce the time-extended $\gamma$-ray emission %
observed at the Sun. This problem is exacerbated by the particularly-hard parent-proton spectrum inferred for the LDGRF, and the recorded photon energies extending to the GeV range. 
These inconsistencies strongly suggest that
the high-energy $\gamma$-ray emission was not caused by the back-precipitation of the same SEP population measured in situ. This result is in agreement with the conclusions by \citet{ref:DENOLFO2019} and \citet{ref:BRUNO2023}, who compared for 14 LDGRFs the numbers of %
protons %
responsible for the 
$>$100 MeV $\gamma$-ray emission and the numbers of protons %
producing the concomitant SEP events recorded at 1 AU (see Section \ref{s:Introduction}).
They demonstrated that
the high precipitation-fraction %
values %
obtained for some events are highly incompatible with the effects 
of magnetic mirroring which, based on modeling, is expected to prevent an efficient
back-propagation of shock-accelerated particles to the dense solar atmosphere \citep{ref:HUTCHINSON2022}.

\item Further evidence against the CME-shock origin of the 2024 July 16 LDGRF
comes from the lack of any apparent link between the durations and time histories
of the $>100$ MeV $\gamma$-ray and the interplanetary type-II radio emissions. 
Specifically, %
the type-II burst accompanying the CME-driven shock acceleration was weak, confined to metric-DH wavelengths, and disappeared about 5.5 hours prior to the LDGRF end time (see Figure \ref{fig:2}). %
This event is therefore inconsistent with the close statistical relationship
claimed by \citet{ref:GOPALSWAMY2018,ref:GOPALSWAMY2019b}, who reported a 1-to-1 linear correlation between the durations of the two phenomena, as well as an anticorrelation between the LDGRF durations and the DH type-II burst end frequencies.
In fact, in addition to the lack of a significant temporal overlap between the two emissions, the 2024 July 16 
LDGRF duration was over 3.5 times longer than that of the concurrent interplanetary type-II emission, 
apparently reflecting distinct underlying production mechanisms with
different timescales.
Furthermore, since the empirical relationship derived by \citet{ref:GOPALSWAMY2019b} is based on the subsample of strongest events with $<$400 kHz end frequencies (see their figure 2), the final type-II frequency of 1.4 MHz exhibited by the 2024 July 16 event is considerably above the upper bound of their dataset. 
Therefore, not only does this place the corresponding data point well outside the domain of their fit, but the extrapolated LDGRF duration predicted at 1.4 MHz is entirely inconsistent with the actual duration observed.
These findings are in line with the investigation of \citet{ref:BRUNO2023} who instead found no significant correlation between the two phenomena.
Indeed, as discussed in Section \ref{s:Introduction}, a direct link between the $\gamma$-ray and DH type-II emissions is not obvious, as they are produced by different particle populations (ions vs. electrons) in distinct heliospheric regions (solar surface vs. interplanetary shock).
The back-precipitation from shocks at the large heliocentric distances (hundreds of R$_{s}$) 
implied by the final interplanetary type-II emission frequency
is strongly impeded both by the magnetic mirroring and the acceleration efficiency, which rapidly increases and decreases with radial distance, respectively. In particular, protons with the energies (typical of GLEs) needed for producing the LDGRF emission are expected to be accelerated and released by the shock within a few R$_{s}$. 

\end{itemize}

Therefore, these findings challenge the interpretation of the 2024 July 16 LDGRF in terms of the CME-shock paradigm, 
and the underlying hypothesis that
the ions producing the $\gamma$-ray emission and the SEP event measured in interplanetary space are part of the same population of shock-accelerated particles.
The alternative LDGRF-origin scenario associated with particle trapping and acceleration within large %
coronal loops, is instead supported by the presence of giant ($>$1 R$_{s}$) arch-like magnetic structures that emerged over the source region right after the eruption and
persisted until well
after the end of the $\gamma$-ray emission at $\sim$21 UT.
The observation of such loops, 
that are generally difficult to visualize %
due to their faint luminosity,
was favored by the limb location far from the brightness of the solar disk, and by the wider FoV of the SUVI instrument on board recent GOES-R series spacecraft compared to SDO/AIA.
The persistence of EUV emission in the 94 \r{A} channel throughout the LDGRF duration, coupled with the absence of cooling signatures in other EUV bands, strongly suggests sustained heating of the loop plasma. This behavior is inconsistent with standard post-flare cooling and instead points to a continuous energy input process.
2$^{nd}$-order Fermi acceleration provides a physically-plausible mechanism for energizing particles to $>$300 MeV over extended timescales. Turbulent-wave activity -- such as Alfv\'enic or compressive magnetosonic modes -- can stochastically accelerate trapped particles and simultaneously heat the ambient plasma via non-resistive dissipation mechanisms \citep{ref:PETROSIAN2012,ref:VANBALLEGOOIJEN2017,ref:CALLINGHAM2024}, 
even though the detailed pathways of energy transfer to the plasma are not yet fully understood.

\section{Summary and Conclusions}\label{s:Summary and Conclusions}  
The 2024 July 16 LDGRF was a remarkable event in terms of photon energies, requiring the presence of a very energetic ion population interacting at the Sun to generate the $>$100 MeV $\gamma$-ray emission measured by LAT. The inferred parent-proton spectrum was indeed one of the hardest ever inferred for a \textit{Fermi} LDGRF, %
second only to the one characterizing the exceptional event on 2017 September 10. 
However,
the contextual multi-spacecraft observations %
show 
that this LDGRF is an ideal counterexample to the CME paradigm.
The shock-related phenomena accompanying the event were weak, 
and exhibit no apparent link
with the $\gamma$-ray emission. Specifically, the CME was relatively slow ($\sim$600 km/s); %
the interplanetary type-II burst disappeared almost 5.5 hours prior to the end of the LDGRF, and was characterized 
by a spectrum that did not extend to kHz frequencies; the type-III burst was also faint and of short duration, unlike the strong bursts typically accompanying significant SEP events.
In addition, the associated SEP event was weak and limited to low particle energies ($<$ few tens of MeV protons) even at magnetically-connected locations. 
These findings demonstrate that the CME-driven shock was unlikely strong enough to accelerate particles to energies well above the pion-production threshold to generate the measured $\gamma$-ray emission, extending to the GeV range.
Consequently, in addition to the points raised by \citet{ref:BRUNO2023}, they strongly challenge the presumed causal connection between particles accelerated at CME-driven shocks and LDGRFs.
We therefore conclude that the 2024 July 16 LDGRF was generated by an ion population distinct from the SEPs measured in situ. 

An interesting feature of this event is the formation of giant hot, quasi-static coronal loops above the source region that persisted throughout the duration of the $\gamma$-ray emission. 
Though conclusive observational support is not available,
we propose that these large-scale bipolar magnetic structures favor the loop-based origin scenario for the LDGRF. In this framework, seed ions are injected and subsequently trapped within the loops, where they undergo 2$^{nd}$-order Fermi acceleration to reach the requisite high energies. These particles continuously energize the loop plasma and eventually diffuse toward the footpoints, precipitating into the solar photosphere and producing $\gamma$-ray emission. This interpretation aligns with the persistent EUV emission observed in the 94 \r{A} channel -- lasting longer than the $>$100 MeV $\gamma$-ray signal -- and the absence of cooling signatures in lower-temperature EUV bands. The detected loop system thus represents a magnetically-closed environment capable of sustaining particle confinement and stochastic acceleration over several hours, consistent with the extended duration of the LDGRF.

Observation of the system of giant hot coronal arches in this case was facilitated by the location of the flaring region close to the limb -- such loops may be hard to identify against the bright disk of the Sun -- and the availability of the EUV images from the SUVI instrument on board recent GOES-R units, which have a larger field of view with respect to SDO/AIA. Thus it is possible that similar looped structures were present at the time of other LDGRFs but were overlooked. The observations during the 2024 July 16 event suggest that further critical investigation of such large-scale hot coronal loops and their possible association with LDGRFs is warranted.

\begin{acknowledgements}
We thank the anonymous reviewer for their time and effort in reviewing this article, and for providing thoughtful and constructive feedback to improve the quality of the work. We are grateful to O. Chris St. Cyr for useful discussion of the CME dynamics.
A.~B. and I.~G.~R. acknowledge support from the National Aeronautics and Space Administration (NASA) Heliophysics Space Weather Research Program's CLEAR SWx Center of Excellence award 80NSSC23M0191. In addition, A.~B. acknowledges support from NASA under award number 80GSFC24M0006, and I.~G.~R. acknowledges support from the STEREO mission. S.~D. acknowledges support from the UK Science and Technology Facilities Council (STFC, grant  ST/Y002725/1). J.~R. acknowledges support from NASA grant 80NSSC21K1375 and support from the National Science Foundation (NSF), grant 2112441.

The \textit{Fermi} LAT Collaboration acknowledges generous ongoing support from a number of agencies and institutes that have supported both the development and the operation of the LAT as well as scientific data analysis. These include NASA and the Department of Energy in the United States, the Commissariat \`a l'Energie Atomique and the Centre National de la Recherche Scientifique / Institut National de Physique Nucl\'eaire et de Physique des Particules in France, the Agenzia Spaziale Italiana and the Istituto Nazionale di Fisica Nucleare in Italy, the Ministry of Education, Culture, Sports, Science and Technology (MEXT), High Energy Accelerator Research Organization (KEK) and Japan Aerospace Exploration Agency (JAXA) in Japan, and the K.~A.~Wallenberg Foundation, the Swedish Research Council and the Swedish National Space Board in Sweden. Additional support for science analysis during the operations phase is gratefully acknowledged from the Istituto Nazionale di Astrofisica in Italy and the Centre National d'\'Etudes Spatiales in France.  MPR and NO are members of the LAT collaboration. 
\end{acknowledgements}

\bibliographystyle{aa}
\bibliography{bibliography}

\end{document}